\documentclass{icrc29}
\usepackage{graphicx,amssymb,amsmath,times}
\begin{document}
%Title of paper
\title[First Estimate of Auger Spectrum]{First Estimate 
of the Primary Cosmic Ray Energy Spectrum\\ 
above 3 EeV from the Pierre Auger Observatory}
\author[Auger Collaboration] {The Pierre Auger Collaboration}
\presenter{Presenter: P. Sommers (sommers@physics.utah.edu)}

\maketitle
%\linenumbers

\begin{abstract}

Measurements of air showers are accumulating at an increasing
rate while construction proceeds at the Pierre Auger Observatory.
Although the southern site is only half complete, the cumulative
exposure is already similar to those achieved by the largest
forerunner experiments.  A measurement of the cosmic ray energy
spectrum in the southern sky is reported here.  The methods
are simple and robust, exploiting the combination of fluorescence
detector (FD) and surface detector (SD).  The methods do not rely
on detailed numerical simulation or any assumption about the
chemical composition.

\end{abstract}

\noindent{\bf \underline{Introduction}}

The southern site of the Pierre Auger Cosmic Ray Observatory in
Argentina now covers an area of approximately
1500 km$^2$.  
On good-weather nights, air fluorescence telescopes record the
longitudinal profiles of extensive air showers in the atmosphere
above the surface array of water Cherenkov detectors
\cite{EApaper,Hybrid}.  Hybrid air shower measurements (FD and SD
together) are utilized in this analysis to avoid dependence on
specific numerical simulations of air showers and detector
responses to them.  The analysis is also free of assumptions
about the primary nuclear masses.  The fluorescence detector (FD)
provides a nearly calorimetric, model-independent energy
measurement: fluorescence light is produced in proportion to
energy dissipation by a shower in the atmosphere \cite{FD}.
Hybrid data establish the relation of shower energy to the ground
parameter S(1000), which is the water Cherenkov signal in the SD
at a distance of 1000 meters from the shower axis.  Moreover,
hybrid data determine the trigger probability for individual
tanks as a function of core distance and energy, from which it
is found that the SD event trigger is fully efficient above 3 EeV
for zenith angles less than 60$^{\circ}$.  The SD exposure is
then calculated simply by integrating the geometric aperture over
time.

\begin{figure}[t]
\begin{minipage}[t]{7.5cm}
\includegraphics*[width=1.0\textwidth]{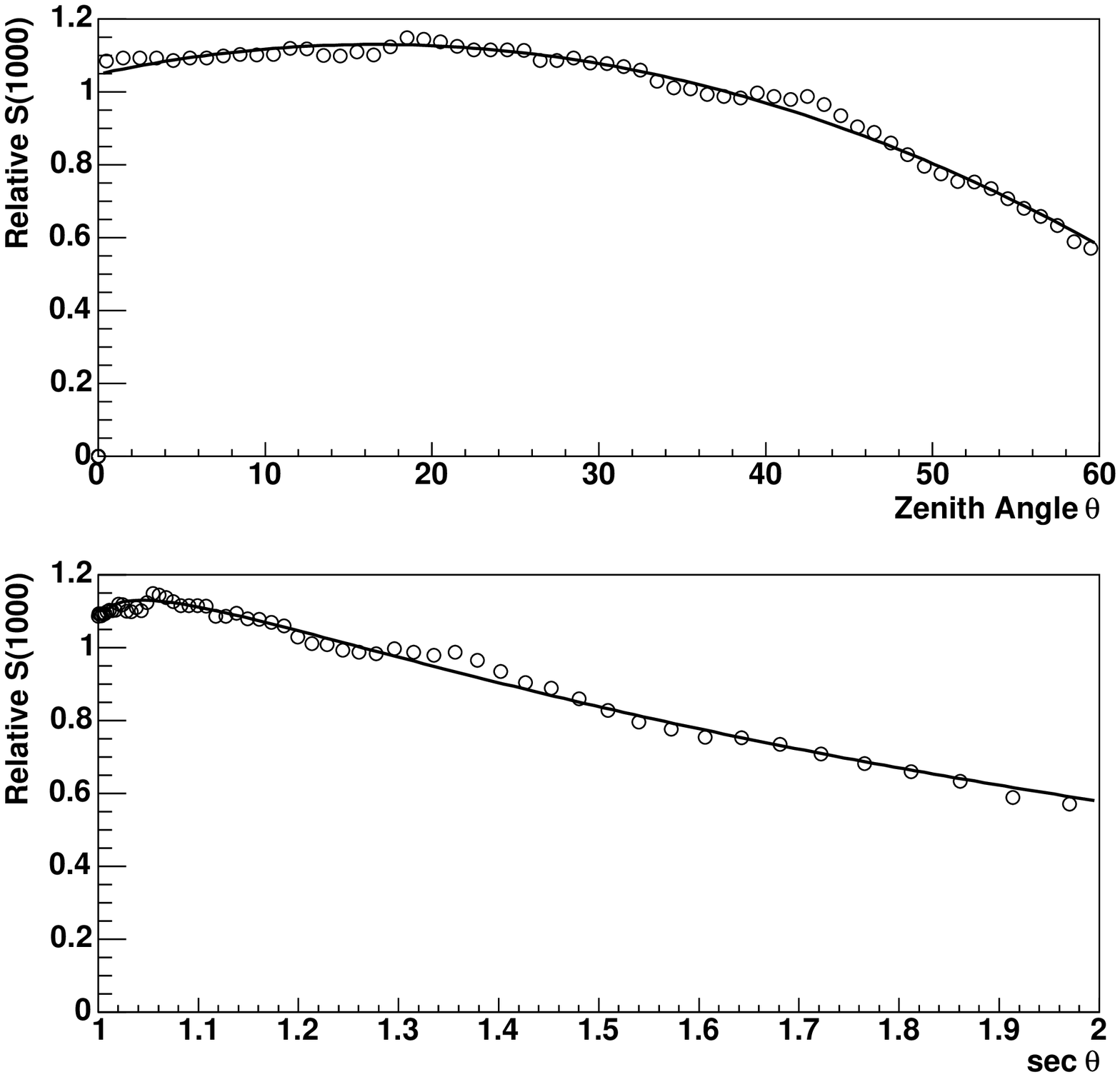}
\caption{\label{f:firstone} The constant intensity curve
$CIC(\theta)$ is determined
  by having the same number of events for $sin\theta cos\theta
  \Delta\theta=0.1$ at each $\theta$ (plotted points are not independent). Values are relative to S(1000) at the median zenith angle of $38^{\circ}$. 
The approximating
quadratic curve is $CIC(\theta)=1.049+0.00974\theta-0.00029\theta^2$.
In the lower plot, the $CIC$ curve
  is replotted as a function of $sec\theta$ to
  exhibit the attenuation of S(1000) with atmospheric slant depth
  at fixed energy.}
\end{minipage}
\hfill
\begin{minipage}[t]{7.5cm}
\includegraphics*[width=1.0\textwidth]{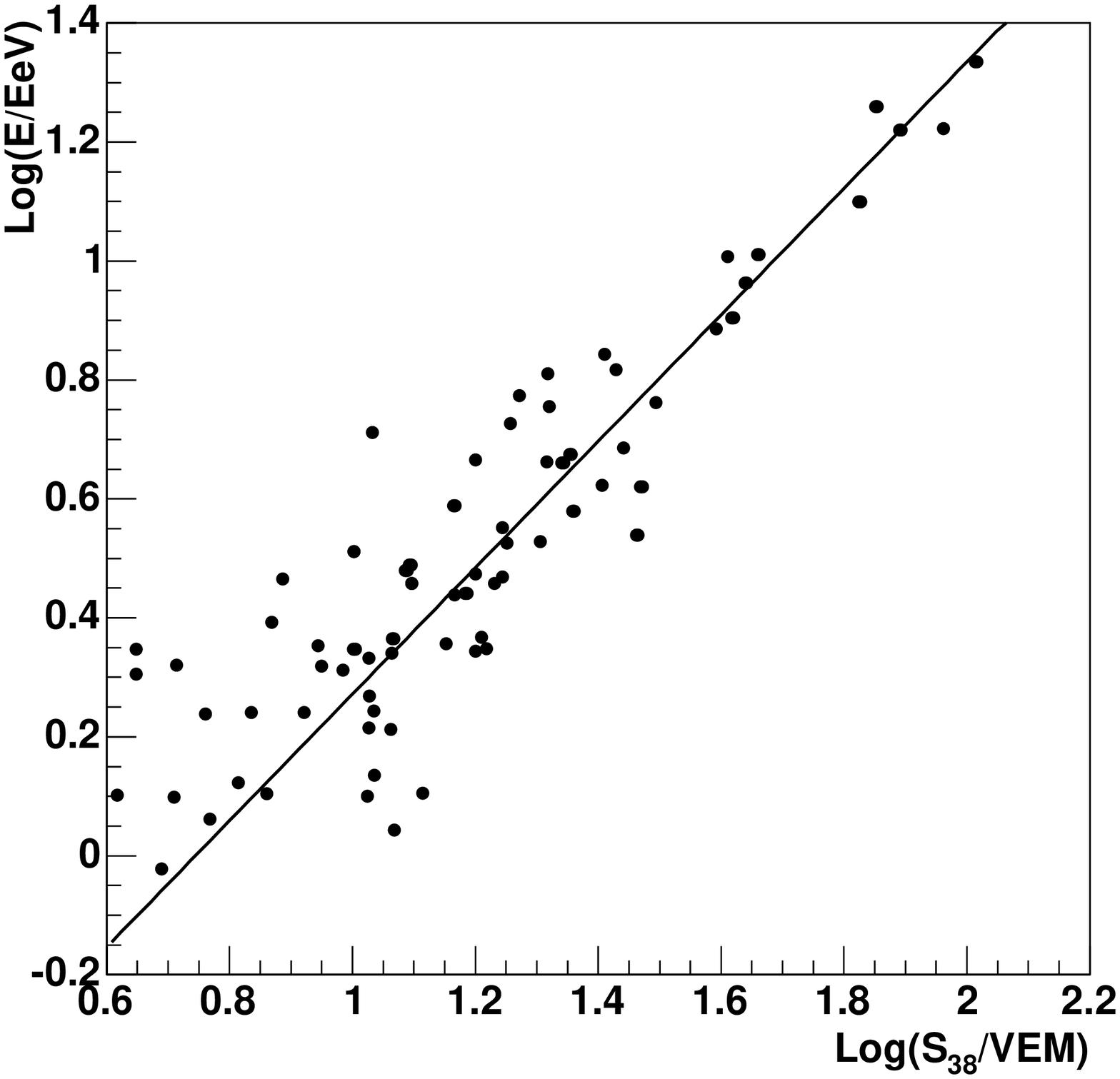}
\caption{\label{f:secondone} FD energy {\em vs.} ground parameter
  $S_{38}$.  These are hybrid events that were recorded when
  there were contemporaneous aerosol measurements, whose FD
  longitudinal profiles include shower maximum in a measured
  range of at least 350 g cm$^{-2}$, and in which there is less than 10\%
  Cherenkov contamination.  The fitted line is
  $Log(E)=-0.79+1.06 Log(S_{38}$).}
\end{minipage}
\hfill
\end{figure}

\begin{figure}[t]
\begin{minipage}[t]{7.5cm}
\includegraphics*[width=1.0\textwidth]{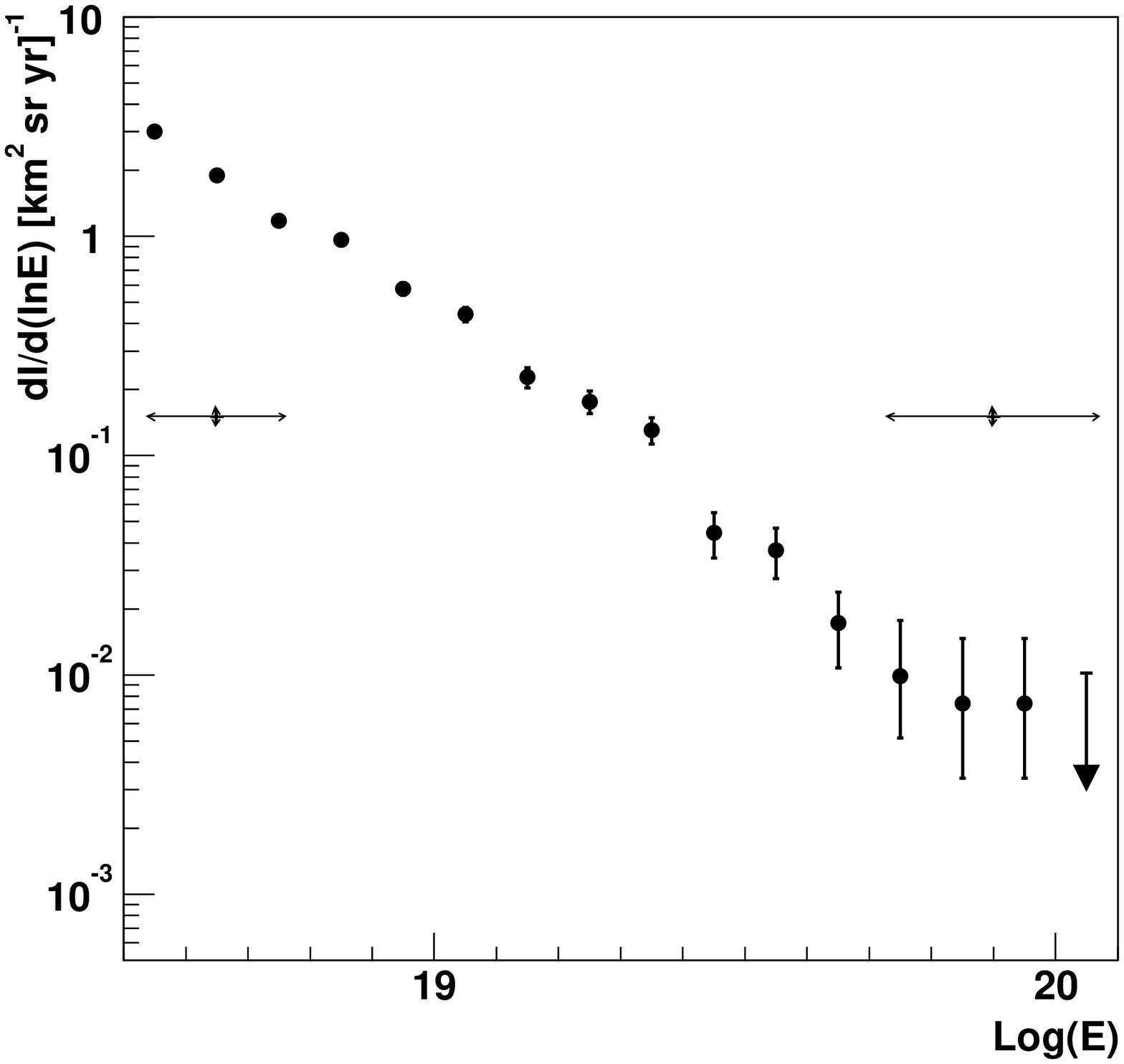}
\caption{\label{f:thirdone} Estimated spectrum.  Plotted on the vertical axis is
the differential intensity $\frac{dI}{dlnE}\equiv
E\frac{dI}{dE}.$  Error bars
  on points indicate statistical
uncertainty (or 95\% CL upper limit).  Systematic uncertainty is indicated by double
arrows at two different energies.}
\end{minipage}
\hfill
\begin{minipage}[t]{7.5cm}
\includegraphics*[width=1.0\textwidth]{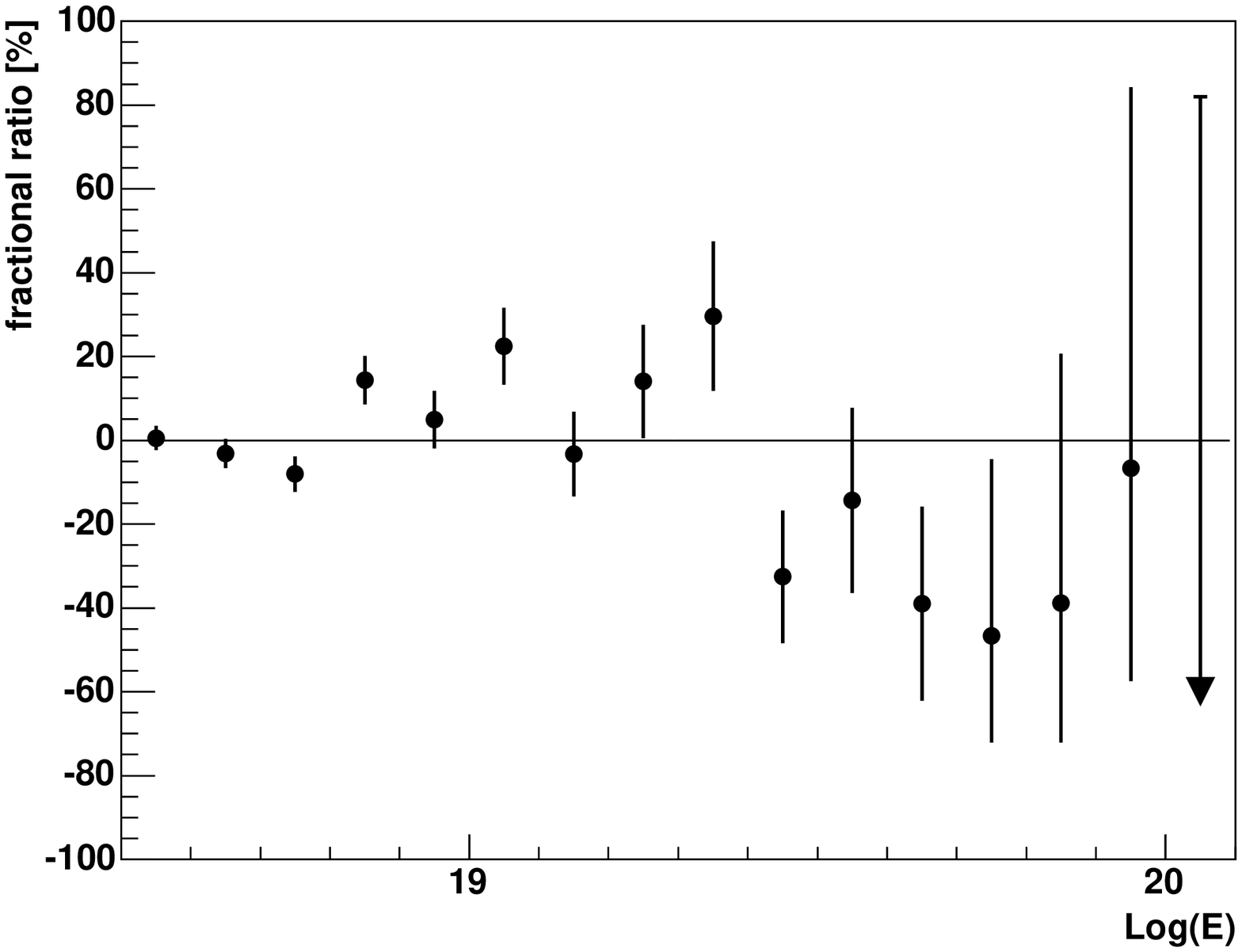}
\caption{\label{f:fourthone} Percentage deviation from the best-fit 
power law: $100\times((dI/d(lnE)-F)/F$.  The fitted function is $F =
30.9\pm 1.7\times(E/EeV)^{-1.84\pm 0.03}$.  The chisquare per degree of freedom
in the fit is 2.4}
\end{minipage}
\hfill
\end{figure}

It is the continuously operating surface array which provides the high
statistics with unambiguous exposure.

The methods adopted for this first analysis are chosen to be
robust and simple.  No event-by-event estimation of shower
penetration is attempted, although a variety of methods to
achieve that may improve the energy resolution in future reports.
The rapidly growing cumulative exposure will provide much higher
statistics for future measurements of the spectrum.  Besides the
present statistical uncertainties, the presentation here also
takes account of unresolved systematic uncertainties.

\noindent{\bf \underline{Analysis methods and results}}

The data for this analysis are from 1 Jan 2004 through 5 Jun 2005.
The event acceptance criteria and exposure calculation are described
in separate papers \cite{Triggers,Acceptance}.  Events are included
for zenith angles 0-60$^{\circ}$, and results are reported for
energies above 3 EeV (3525 events). The array is fully efficient for
detecting such showers, so the acceptance at any time is the simple
geometric aperture.  The cumulative exposure adds up to 1750 km$^2$ sr
yr, which is 7\% greater than the total exposure obtained by AGASA
\cite{AGASA}.  The average array size during the time of this exposure
was 22\% of what will be available when the southern site of the
Observatory has been completed.

Assigning energies to the SD event set is a two-step process.
The first step is to assign an energy parameter $S_{38}$ to each
event.  Then the hybrid events are used to establish
the rule for converting $S_{38}$ to energy.

The energy parameter $S_{38}$ for each shower comes from its
experimentally measured S(1000), which is the time-integrated
water Cherenkov signal S(1000) that would be measured by a tank
1000 meters from the core.  This ground parameter is determined
accurately by non-linear interpolation even when there is no tank
at that particular core distance \cite{LDF}.  

The slant depth of the surface array varies from 870 g cm$^{-2}$
for vertical showers to 1740 g cm$^{-2}$ for showers at zenith
angle $\theta=60^{\circ}$.  The signal S(1000) is
attenuated at large slant depths.  Its dependence on zenith angle
is derived empirically by exploiting the nearly isotropic
intensity of cosmic rays.  By fixing a specific intensity $I_0$
(counts per unit of $sin^2\theta$), one finds for each zenith
angle the value of S(1000) such that $I(>S(1000))=I_0$.  A
particular constant intensity cut gives the curve $CIC(\theta)$
of figure 1.  The S(1000) values are shown relative to the value
at the median zenith angle ($\theta\approx 38^{\circ}$).  Given
S(1000) and $\theta$ for any measured shower, the energy
parameter $S_{38}$ is defined by {\bf $S_{38}\equiv
S(1000)/CIC(\theta)$}.  It may be regarded as the S(1000)
measurement the shower would have produced if it had arrived
$38^{\circ}$ from the zenith.  

This formula for $S_{38}$ implicitly assumes that all constant
intensity curves are simple rescalings of the reference curve
$CIC(\theta)$ of figure 1, which corresponds to $S_{38}=15$ VEM
(vertical equivalent muons).  Higher values of $S_{38}$ also yield
curves of constant intensity to the accuracy that can be checked with
current statistics.  With a much larger data set, it will be possible
to investigate any change in shape of the constant intensity curve
with energy and thereby reduce any systematic error that might be
associated with this simple formula for $S_{38}$.

$S_{38}$ is well correlated with the FD energy measurements in
hybrid events that are reconstructed independently by the FD and
SD.  See figure 2.  The fitted line gives an empirical rule for
assigning energies (in EeV) based on $S_{38}$ (in VEM):
\begin{equation}E = 0.16 \times S_{38}^{1.06} = 
  0.16\times [S(1000)/CIC(\theta)]^{1.06}.\end{equation}
The uncertainty in this rule is discussed below.  The hybrid
events in figure 2 start at $\sim 1$ EeV.  The acceptance is not
saturated below 3 EeV, but the events used in figure 2 are those
with core locations and arrival directions such that they have
probability greater than 0.9 for satisfying the SD trigger and
quality conditions.  These events increase the statistics and the
``moment arm'' of the correlation without introducing appreciable
bias.

The distribution over $ln(E)$ produced by this two-step procedure
becomes the energy spectrum of figures 3 and 4 after dividing by the
exposure: 1750 km$^2$ sr yr.  (See also
http://www.auger.org/icrc2005/spectrum.html.)

\noindent{\bf \underline{Uncertainties and caveats}}

The Auger Observatory will measure the spectrum over the southern
sky accurately in coming years.  The spectrum in figure 3 is only
a first estimate.  It has significant systematic and statistical
uncertainties.  The indicated statistical error for each point
comes directly from the Poisson uncertainty in the number of
measured showers in that logarithmic energy bin.  Systematic and
statistical uncertainties in S(1000) are discussed elsewhere
\cite {Errors}.  There is larger systematic uncertainty in the
conversion of $S_{38}$ to energy.  Part of that comes from the FD
energies themselves.  Laboratory measurements of the fluorescence
yield are uncertain by 15\%, and the absolute calibration of the
FD telescopes is presently uncertain by 12\%.  Together with
other smaller FD uncertainties, the total systematic uncertainty
in the FD energy measurements is estimated to be 25\%.  Another
part of the systematic energy uncertainty in this analysis comes
from quantifying the correlation in figure 2.  The accuracy is
limited by the available statistics, and the uncertainty grows
with energy.  Combining in quadrature the FD systematic
uncertainty and this correlation uncertainty, the total
systematic energy uncertainty grows from 30\% at 3 EeV to 50\% at
100 EeV.  This uncertainty is indicated by horizontal double
arrows in figure 3, and a 10\% systematic uncertainty in the
exposure is indicated by vertical arrows.

The fraction of primary energy that does not contribute to
fluorescence light (due to neutrinos, muons, and other weakly
interacting particles) has an estimated uncertainty of 4\% due to
the unknown primary mass and differences in viable hadronic
interaction models.  This is included in the 25\% uncertainty for
FD energies.  It should be acknowledged, however, that it is not
possible empirically to rule out larger amounts of ``missing
energy'' due to exotic particle physics.  In principle, some
shower energies could be underestimated.

The spectrum of figure 3 is based on the water Cherenkov signal
S(1000).  Primary photons would be expected to produce
a smaller S(1000) signal due to the lack of muons.  On average,
the S(1000) from a photon primary should be roughly one-half the
signal for a hadronic primary of the same energy.  The Auger
Observatory will eventually obtain upper limits on the photon
flux at all energies.  For now, the limit \cite{Photons} does not
pertain to the highest energies, so the results in figure 3 in
the highest energy bins are predicated on the assumption that
primary photons are not a major component.

\noindent{\bf \underline{Discussion and a look to the future}}

The Pierre Auger Observatory is still under construction and
growing rapidly.  By the next ICRC meeting, its cumulative
exposure will be approximately 7 times greater.  The statistical
errors will shrink accordingly, permitting a search in the
southern skies for spectral features, including the predicted GZK
suppression.  The enlarged hybrid data set will reduce systematic
uncertainty in the FD normalization of the SD energies.

Numerous laboratory experiments are attempting to reduce the
systematic uncertainty in the fluorescence yield, which will 
be the dominant uncertainty in the FD normalization of the
Auger energy spectrum.  The FD detector calibration uncertainty
will also be reduced.

Preliminary studies based on comparing real data with simulation data
give energies that are systematically higher than the FD-normalized
energies by approximately 25\%.  This number has some dependence on
the hadronic model, the primary mass and the shower propagation code
that are assumed.  The Pierre Auger Observatory is uniquely configured
for the investigation of this intriguing difference.  Measurements of
$\mathrm{X_{max}}$, LDF steepness, signal rise time, and shower front curvature
complement the measurement of S(1000).  Studying the distributions of
these shower properties at different zenith angles and energies can
constrain the cosmic ray composition and hadronic interaction models.
The parameters might be used shower-by-shower to improve the accuracy
of energy determinations.  Future measurements of the energy spectrum
will be based on vastly larger data sets and may also exploit the rich
information that is available for each shower.


\vspace{-7mm}
\begin{thebibliography}{000}
\bibitem{AGASA}M. Takeda et al., {Astroparticle Physics} {\bf
  19}, 447-462 (2003) 
\bibitem{EApaper}Auger Collaboration, {NIM} {\bf 523}, 50 (2004)
\bibitem{Hybrid} ``Hybrid performance of the Pierre Auger Observatory,'' usa-mostafa-M-abs1-he14-oral
\bibitem{FD} ``Performance of the fluorescence detectors of the Pierre
  Auger Observatory,'' aus-bellido-J-abs1-he14-oral
\bibitem{Triggers} ``The trigger system of  the PAO surface detector:
  operation, efficiency and  stability,'' usa-bauleo-PM-abs1-he14-poster
\bibitem{Acceptance} ``Determination of the aperture of the PAO surface detector,'' fra-parizot-E-abs1-he14-poster
\bibitem{LDF} ``Measurement of the lateral distribution function of UHECR air showers,'' usa-bauleo-PM-abs2-he14-poster
\bibitem{Errors} ``Statistical and systematic uncertainties in the
  event reconstruction and S(1000) determination by the Pierre
  Auger surface detector,'' ita-ghia-P-abs1-he14-oral
\bibitem{Photons} ``Upper limit on the primary photon fraction
  from the Pierre Auger Observatory,'' ger-risse-M-abs2-he14-oral
\end{thebibliography}
\end{document}